\documentclass[11pt]{article}

\usepackage[margin=1in]{geometry}
\usepackage[T1]{fontenc}
\usepackage{amsmath,amssymb}
\usepackage{graphicx}
\usepackage{booktabs}
\usepackage[round,authoryear]{natbib}
\usepackage{bm}
\usepackage{tabularx}
\usepackage{array}
\newcolumntype{L}{>{\raggedright\arraybackslash}X}
\newcolumntype{R}{>{\raggedleft\arraybackslash}X}
\newcommand{\sep}{\unskip;\space}
\newenvironment{keyword}{\par\smallskip\noindent\textbf{Keywords: }\ignorespaces}{\par}

\graphicspath{{figures/}}

\usepackage[hidelinks]{hyperref}

\begin{document}

\title{Same-Hour Estimation and Multi-Horizon Forecasting of PM$_{2.5}$ in Beijing: Model Comparison and Feature-Group Contributions}

\author{Yusheng Wang\thanks{Corresponding author: wangyusheng135@gmail.com} \and Ruiyan Du\\[0.4em]
\small Northeastern University at Qinhuangdao, Qinhuangdao 066004, China}
\date{}
\maketitle

\begin{abstract}
Accurate PM$_{2.5}$ prediction depends on information available when a prediction is issued.
Using 420{,}768 hourly observations from 12 monitoring stations in Beijing between March 2013
and February 2017, we compare same-hour estimation ($h=0$) with strict one-, six-, and
24-hour-ahead forecasting.
Stations are reindexed to complete hourly calendars before lagged and rolling features are
constructed.
Chronological train, calibration, and test blocks and expanding-window preprocessing prevent
target-time leakage.
Persistence and seasonal-naive baselines are compared with Ridge, Elastic Net, and a five-seed
XGBoost ensemble.
Ridge attains RMSE 20.15 at one hour, XGBoost 54.13 at six hours, and Elastic Net 80.83 at
24 hours, improving on persistence by 5.7\%, 11.8\%, and 17.6\%, respectively.
Same-hour XGBoost reaches RMSE 14.03, but feature ablations attribute most of this performance
to recent PM$_{2.5}$ history and synchronous PM$_{10}$.
Moving-block intervals, repeated seeds, and leave-one-station-out evaluation quantify uncertainty
and spatial transfer.
The results support sensor cross-checking at $h=0$, routine updates at $h=1$, alert preparation
at $h=6$, and cautious daily planning at $h=24$.
\end{abstract}


\section{Introduction}

Fine particulate matter (PM$_{2.5}$) is a major indicator of urban air pollution and is
associated with respiratory and cardiovascular risks as well as reduced atmospheric visibility.
Short-horizon concentration estimates can support routine monitoring, public warnings,
traffic management, and the allocation of emission-control resources.
Beijing is an informative setting because concentrations vary substantially across seasons,
locations, and weather regimes, while its long-running monitoring network makes it possible to
examine both temporal dependence and station heterogeneity.
Severe pollution episodes and later air-quality improvements in the city have also been
documented in environmental studies~\citep{zhang2017cautionary}.

Air-quality prediction has developed from regression and deterministic dispersion models toward
machine-learning approaches that combine pollutant, meteorological, temporal, and spatial
predictors.
Reviews of this literature describe a broad set of linear, tree-based, neural, and hybrid models,
but also emphasize that reported accuracy depends strongly on the pollutant, prediction horizon,
input availability, and evaluation design~\citep{rybarczyk2018systematic,cabaneros2019review,masood2021review}.
Recent reviews further organize the field around deep-learning, graph-based, hybrid, and
spatiotemporal approaches, while emphasizing that information availability and evaluation design
determine whether reported gains are operationally comparable~\citep{korzeniewska2026,agbehadji2024,zhou2024}.
Regularized regression remains useful when explanatory variables are highly correlated:
PM$_{2.5}$, PM$_{10}$, CO, and NO$_{2}$ often reflect related emission and dispersion conditions.
Ridge, Lasso, and Elastic Net stabilize or select coefficients under this multicollinearity
~\citep{hoerl1970ridge,tibshirani1996lasso,zou2005elasticnet}, while tree ensembles such as
XGBoost can represent nonlinear interactions and threshold effects~\citep{chen2016xgboost}.

Recent studies have shown that recurrent, convolutional, and graph-based models can capture
temporal and spatiotemporal structure in air-pollution series~\citep{li2017lstm,huang2018cnn,qi2019hybrid}.
These approaches are valuable comparators, especially where long histories or network structure
are available.
The present study is not intended as an exhaustive benchmark of architectures such as STGCN,
PatchTST, or Informer.
Instead, it establishes interpretable tabular reference bounds under a common
information-availability protocol, without introducing neural-network window padding, masking,
or graph-topology assumptions that would change the comparison.
Their reported scores, however, are not directly interchangeable: a result obtained with a
contemporaneous pollutant measurement, a random split, or a short lead time answers a different
operational question from a forecast issued before the target hour.
A clear distinction between the information origin and the target time is therefore as important
as the choice of estimator.

This distinction is also relevant to monitoring-system design.
Same-hour estimates based on co-pollutants can provide useful redundancy when a target channel
is delayed or requires quality control, complementing rather than replacing direct measurements.
Low-cost and auxiliary sensing systems are increasingly used for this purpose, but their practical
value depends on calibration, environmental sensitivity, and transparent communication of
uncertainty~\citep{kumar2015rise,morawska2018applications,karagulian2019review,rai2017enduser,malings2019calibration,barkjohn2021purpleair}.
A same-hour PM$_{2.5}$ estimate should consequently be interpreted as a sensor-estimation or
cross-checking task, whereas a forecast must be based only on information available before the
target time.

Evaluation choices can otherwise make time-series prediction appear more accurate than it would
be in operation.
Randomly mixing observations from different dates can allow training data to resemble, or
indirectly inform, the test period.
Forecasting studies therefore commonly use chronological holdouts and expanding or rolling
validation windows~\citep{tashman2000outofsample,bergmeir2012crossvalidation,cerqueira2020evaluating}.
With multiple monitoring stations, the split should also respect the spatial and temporal
structure of the data rather than treating station-hour records as independent draws
~\citep{roberts2017crossvalidation}.

The same concern applies to feature interpretation.
A synchronous pollutant can be highly informative for reconstructing a target measurement while
offering little advance warning of a future target.
Likewise, a lagged target can dominate a one-hour forecast even when a large feature matrix
is available.
Feature importance or a high $R^2$ is therefore not, by itself, evidence that a model has learned
a transferable pollution mechanism.
It is more useful to report the information set, the forecast horizon, and the change relative to
an operational baseline together.
This perspective also makes comparisons across model families more transparent: a flexible model
should be credited only for improvement that survives the same temporal split and uncertainty
calculation.

Using the Beijing Multi-Site Air-Quality Data Set~\citep{ucibeijing2017}, this study compares one
same-hour estimation task with strict one-, six-, and 24-hour-ahead forecasts in a common
framework.
The analysis combines simple, interpretable regularized models with a nonlinear XGBoost ensemble;
uses chronological training, calibration, and test blocks; and reports paired moving-block
uncertainty, repeated-seed variation, and leave-one-station-out evaluation.
The design makes it possible to separate two complementary questions: how much information is
available for an immediate sensor estimate, and how model performance changes when a real
forecasting lead time is required.

The study addresses three questions:
\begin{enumerate}
    \item How do prediction errors and model rankings change from same-hour estimation to
          one-, six-, and 24-hour-ahead forecasting?
    \item How much information is contributed by recent PM$_{2.5}$ history, synchronous
          co-pollutants, meteorology, temporal variables, and station indicators?
    \item How stable are the principal comparisons under temporal resampling, repeated random
          seeds, and evaluation at monitoring stations excluded from model training?
\end{enumerate}

The remainder of the paper is organized as follows.
Section~\ref{sec:methods} introduces the models, metrics, and chronological evaluation.
Section~\ref{sec:data} describes the data, prediction tasks, and feature construction.
Section~\ref{sec:results} presents the multi-horizon results and ablations.
Section~\ref{sec:discussion} discusses the monitoring implications and scope of the findings.
Section~\ref{sec:conclusion} concludes the paper.

\section{Materials and Methods}
\label{sec:methods}

\subsection{Regularized and Nonlinear Models}

Let the data contain $n$ samples, with PM$_{2.5}$ response $y_i\in\mathbb{R}$ and feature vector
$\mathbf{x}_i\in\mathbb{R}^p$.
Ridge regression adds an $L_2$ penalty to stabilize coefficients under multicollinearity.
Elastic Net combines $L_1$ and $L_2$ penalties:
\begin{equation}
\min_{\beta_0,\boldsymbol{\beta}}
\left\{
\frac{1}{2n}\sum_{i=1}^{n}(y_i-\beta_0-\mathbf{x}_i^\top\boldsymbol{\beta})^2
+\lambda\left[\frac{1-\alpha}{2}\lVert\boldsymbol{\beta}\rVert_2^2
+\alpha\lVert\boldsymbol{\beta}\rVert_1\right]
\right\}.
\end{equation}
The Ridge grid contains 11 log-spaced values from $10^{-3}$ to $10^2$.
The Elastic Net grid contains $\lambda\in\{0.01,0.1,1,10\}$ and $\alpha\in\{0.5,0.9,1.0\}$.
The selected value is $\alpha=1.0$ in all four tasks, indicating that sparsity is the dominant
regularization form for this data set.
At this boundary value, the fitted penalty is mathematically equivalent to the Lasso (the
$L_2$ term has zero weight); we retain the name ``Elastic Net'' to describe the registered
hyperparameter search, but label the fitted result as Lasso-equivalent or sparse linear in
the interpretation below.
Coordinate descent is used for the sparse model~\citep{friedman2010regularization}.

The model set was chosen to make the comparison interpretable across horizons.
Ridge provides a stable linear reference when many lagged and pollutant variables are correlated;
Elastic Net tests whether a sparse subset is sufficient; and XGBoost represents nonlinear
threshold and interaction effects without requiring a station sequence to be padded or pooled
into a neural-network window.
This deliberately modest model set provides transparent tabular reference bounds rather than
a claim that these estimators dominate modern deep spatiotemporal architectures.
This selection also allows the results to distinguish a change in predictive information from
a change in model flexibility.

XGBoost is included as a nonlinear alternative.
It uses histogram tree construction, depth 6, learning rate 0.05, row subsampling 0.8,
column subsampling 0.8, and minimum child weight 5.
These moderate settings were fixed before the final comparison to limit learner-specific
degrees of freedom; only the number of trees was tuned within the expanding validation design.
The number of trees is selected by early stopping on the final internal expanding-training fold.
The selected counts were 220, 137, 90, and 69 for $h=0$, 1, 6, and 24, respectively.
This tree count is fixed when the model is refitted on the full training block for random seeds
11, 22, 33, 44, and 55.
The mean of the five predictions is reported as the XGBoost ensemble.
Repeated seeds are reported because a small numerical advantage from a stochastic learner should
be interpreted differently from a horizon-scale change in error.

\subsection{Evaluation Metrics}

Performance is measured by root mean squared error (RMSE), mean absolute error (MAE), and the
coefficient of determination ($R^2$):
\begin{equation}
\mathrm{RMSE}=\sqrt{\frac{1}{n}\sum_{i=1}^{n}(y_i-\hat y_i)^2},\quad
\mathrm{MAE}=\frac{1}{n}\sum_{i=1}^{n}|y_i-\hat y_i|,\quad
R^2=1-\frac{\sum_i(y_i-\hat y_i)^2}{\sum_i(y_i-\bar y)^2}.
\end{equation}

Persistence is the principal operational baseline.
For $h>0$, it predicts the future target with the current PM$_{2.5}$ value $z_{s,t}$; for
same-hour estimation, it uses $z_{s,t-1}$.
A seasonal-naive baseline uses the value at the same clock hour on the previous day whenever it
is available.
Missing baseline values fall back to persistence and then to the training mean.
Reporting improvement relative to persistence at the same horizon prevents a low absolute RMSE
at a short horizon from being mistaken for a large operational gain.

\subsection{Chronological Validation and Uncertainty}

For each prediction horizon, unique target timestamps are divided chronologically into 70\%
training, 15\% calibration, and 15\% final test blocks.
All stations sharing a target timestamp remain in the same partition.
This protects against a situation in which different stations at the same hour appear in both
training and testing.
Four expanding-window folds are formed inside the training block, beginning with at least 40\%
of the training period.
Numerical imputation, missingness handling, and standardization are fitted separately within
each training fold, so later observations do not determine earlier preprocessing parameters.
The scheme follows the out-of-sample logic recommended for forecasting and structured data
~\citep{hyndman2021forecasting,tashman2000outofsample,cerqueira2020evaluating}.

Hourly observations remain serially dependent even after chronological splitting.
Sampling uncertainty is therefore estimated with 500 moving-block bootstrap replicates using
contiguous blocks of 168 target hours.
All stations within the same target timestamp stay together, and identical resampled blocks are
used for a model and persistence when computing paired RMSE improvements.
This paired construction focuses the interval on the operational quantity of interest: whether a
model improves on the available baseline rather than whether two independently resampled RMSE
values differ.

The calibration block also supplies residuals for split-conformal interval diagnostics
~\citep{lei2018distribution}.
Because calibration and test observations in this study are serially dependent and seasonally
nonstationary, standard split-conformal exchangeability is not assumed; the intervals are
empirical diagnostics rather than strict distribution-free guarantees.
Sequential methods such as EnbPI and adaptive conformal inference are promising alternatives
for dependent air-quality streams~\citep{xu2021enbpi,gibbs2021adaptive}.
These intervals are reported as calibration diagnostics rather than as a guarantee under serial
dependence; probabilistic forecasts require both sharpness and reliable coverage
~\citep{gneiting2014probabilistic}.
Spatial transfer is evaluated by holding out each station in turn, removing station indicators,
and fitting pooled Ridge models to the remaining stations.
Holding out entire stations complements chronological testing by asking whether a relation
learned from the network remains useful away from the stations used for fitting.

\subsection{Data and Experimental Setup}
\label{sec:data}

\subsubsection{Dataset}

The Beijing Multi-Site Air-Quality Data Set contains hourly observations from 12 monitoring
stations between 1 March 2013 and 28 February 2017.
The stations are Aotizhongxin, Changping, Dingling, Dongsi, Guanyuan, Gucheng, Huairou,
Nongzhanguan, Shunyi, Tiantan, Wanliu, and Wanshouxigong.
The raw data contain 420{,}768 station-hour records and include PM$_{2.5}$, PM$_{10}$,
SO$_{2}$, NO$_{2}$, CO, O$_{3}$, temperature, pressure, dew point, rainfall, wind speed
(WSPM, m~s$^{-1}$), wind direction, and calendar fields.
This combination allows the analysis to represent local persistence, co-pollutant association,
weather conditions, and recurrent calendar effects without importing future observations.

PM$_{2.5}$ is missing in 8{,}739 records (2.08\%); rows without a target value are excluded from
supervised evaluation.
Other predictors can still be unavailable at a particular station-hour.
Rather than deleting those hours and changing the temporal sample, the preprocessing pipeline
retains the record and uses a median imputer fitted only on the training portion of each
expanding fold (and refitted on the complete training block for the final model).
A binary missingness indicator is retained for every lagged/rolling PM$_{2.5}$ feature,
co-pollutant, meteorological variable, and wind-direction input.
This keeps the evaluation focused on the information that would be available at the forecast origin.

\subsubsection{Information Availability and Prediction Tasks}

\begin{figure}[htbp]
    \centering
    \includegraphics[width=0.92\textwidth]{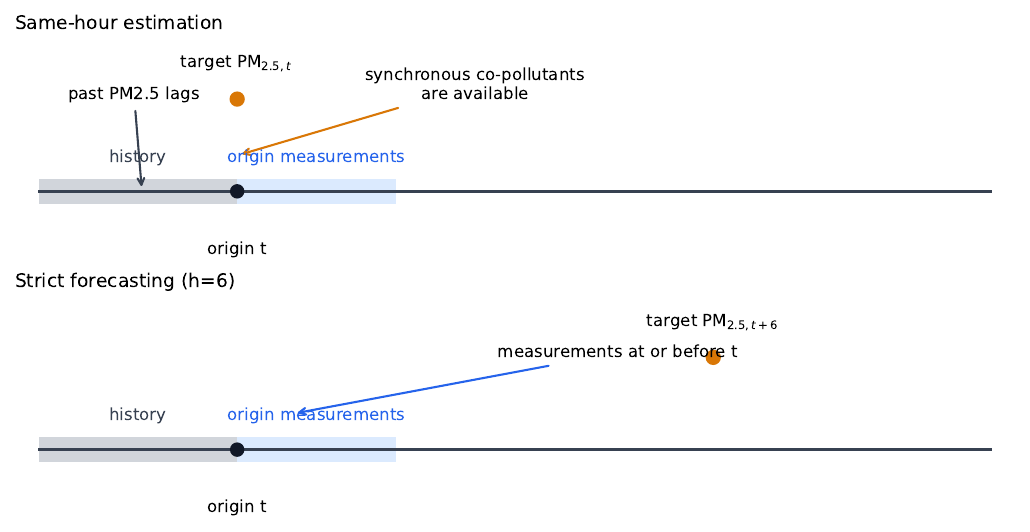}
    \caption{Information available for same-hour estimation and strict forecasting.
    Measurements at or before origin $t$ are valid forecast inputs, whereas a future target
    is never included among the predictors.}
    \label{fig:timeline}
\end{figure}

Let $s$ index a station and $z_{s,t}$ denote PM$_{2.5}$ at time $t$.
The strict forecasting targets are
\begin{equation}
y_{s,t}^{(h)}=z_{s,t+h},\qquad h\in\{1,6,24\}.
\end{equation}
Every feature associated with origin $(s,t)$ is available at time $t$ or earlier.
The current PM$_{2.5}$ value is therefore a valid forecast input for $h>0$.
For the $h=0$ same-hour task, the current target is excluded from the feature vector, while
preceding PM$_{2.5}$ lags and synchronous co-pollutants are retained.
The latter measurements are contemporaneous with the target and are consequently valid for a
sensor-estimation task but not evidence of future predictive lead time.
Figure~\ref{fig:timeline} summarizes the two information settings.

\subsubsection{Preprocessing and Feature Engineering}

Each station is first reindexed to a complete hourly calendar.
Lags and rolling windows are then generated separately by station, preventing a missing record
or station boundary from changing the clock time represented by a feature.
PM$_{2.5}$ lags are constructed at 1, 3, 6, 12, 24, 48, and 72 hours.
Rolling means use the preceding 6 and 24 hours, with the shift applied before aggregation so
that the current target cannot enter a rolling feature.

Origin-time covariates include five co-pollutants, five meteorological variables, wind direction,
station indicators, and calendar features.
Hour and month are also represented by sine and cosine terms.
Each lagged or rolling feature has a missingness indicator, while numerical values are imputed
within the model pipeline.
Tracking variables used for alignment and partitioning are excluded from the model matrix.
The full design contains 81 features for $h=0$ and 82 features for $h>0$, where the additional
forecast feature is current PM$_{2.5}$.
Meteorological inputs are observed origin-time values; this matches retrospective monitoring
analysis, while an operational 24-hour deployment would substitute weather forecasts for those
inputs.

The feature construction is performed in a fixed order.
First, each station is aligned to the complete hourly calendar.
Second, target timestamps are shifted to form the four prediction tasks.
Third, lags, rolling summaries, and their missingness indicators are created within station.
Finally, the chronological partitions are applied and the model pipelines are fitted only on the
relevant training folds.
This order matters because filling a gap before station alignment can change the meaning of a
lag, while calculating a rolling mean before shifting can allow the target hour to enter its own
predictor.
The resulting design is deliberately traceable: every predictor can be mapped to a station,
an origin timestamp, and an information-availability rule.

Hyperparameters are selected inside the expanding training folds rather than from the final
test block.
The selected preprocessing parameters and model settings are then refitted on the complete
training block before calibration and final testing.
For XGBoost, early stopping determines the tree count once, and that count is reused across
the five final seeds.
This procedure separates model selection from the final comparison and makes the reported seed
variation a property of the fitted learner rather than a mixture of tuning and testing variability.

\section{Results}
\label{sec:results}

\subsection{Data Characteristics}

Table~\ref{tab:descriptive} summarizes the principal variables.
PM$_{2.5}$ has a mean of 79.79~$\mu$g/m$^3$ and a median of 55.00~$\mu$g/m$^3$, reflecting a
strongly right-skewed concentration distribution.
The spread between the median and the 95th percentile (242.00~$\mu$g/m$^3$) means that a model
is evaluated across both ordinary and high-concentration hours rather than around a single
typical level.

Table~\ref{tab:correlations} reports pairwise Spearman rank correlations among the principal
variables.
PM$_{2.5}$ shows strong positive rank correlation with PM$_{10}$ ($\rho=0.891$),
CO ($\rho=0.837$), and NO$_{2}$ ($\rho=0.658$), while WSPM shows a moderate negative rank
correlation ($\rho=-0.329$).
These associations characterize the modeling problem by identifying correlated and proxy
predictor groups, motivating the use of regularized models and framing the feature-group
ablations reported below.
The coefficients are computed over available numeric raw-record pairs and represent descriptive
associations rather than causal effects.

\begin{table}[htbp]
\caption{Descriptive statistics of the raw station records. WSPM denotes wind speed (m~s$^{-1}$).}
\label{tab:descriptive}
\small
\begin{tabularx}{\textwidth}{LRRRRR}
\toprule
\textbf{Variable} & \textbf{Mean} & \textbf{SD} & \textbf{Median} &
\textbf{95th pct.} & \textbf{Max}\\
\midrule
PM$_{2.5}$ ($\mu$g m$^{-3}$) & 79.793 & 80.822 & 55.000 & 242.000 & 999.000\\
PM$_{10}$ ($\mu$g m$^{-3}$) & 104.603 & 91.772 & 82.000 & 279.000 & 999.000\\
CO ($\mu$g m$^{-3}$) & 1230.766 & 1160.183 & 900.000 & 3500.000 & 10000.000\\
NO$_{2}$ ($\mu$g m$^{-3}$) & 50.639 & 35.128 & 43.000 & 117.000 & 290.000\\
TEMP ($^{\circ}$C) & 13.539 & 11.436 & 14.500 & 30.600 & 41.600\\
WSPM (m s$^{-1}$) & 1.730 & 1.246 & 1.400 & 4.300 & 13.200\\
\bottomrule
\end{tabularx}
\end{table}

\begin{table}[htbp]
\caption{Pairwise Spearman rank-correlation matrix for the principal pollutant and
meteorological variables. Values are computed over available numeric raw-record pairs;
they describe statistical associations and are not causal effects. WSPM denotes wind speed
(m~s$^{-1}$).}
\label{tab:correlations}
\small
\begin{tabular}{lrrrrrr}
\toprule
 & PM$_{2.5}$ & PM$_{10}$ & CO & NO$_{2}$ & TEMP & WSPM\\
\midrule
PM$_{2.5}$ & 1.000 & 0.891 & 0.837 & 0.658 & -0.024 & -0.329\\
PM$_{10}$  & 0.891 & 1.000 & 0.732 & 0.643 & -0.018 & -0.251\\
CO         & 0.837 & 0.732 & 1.000 & 0.741 & -0.222 & -0.389\\
NO$_{2}$     & 0.658 & 0.643 & 0.741 & 1.000 & -0.270 & -0.441\\
TEMP       & -0.024 & -0.018 & -0.222 & -0.270 & 1.000 & 0.102\\
WSPM       & -0.329 & -0.251 & -0.389 & -0.441 & 0.102 & 1.000\\
\bottomrule
\end{tabular}
\end{table}

\subsection{Model Performance Across Forecast Horizons}

\begin{figure}[htbp]
    \centering
    \includegraphics[width=0.86\textwidth]{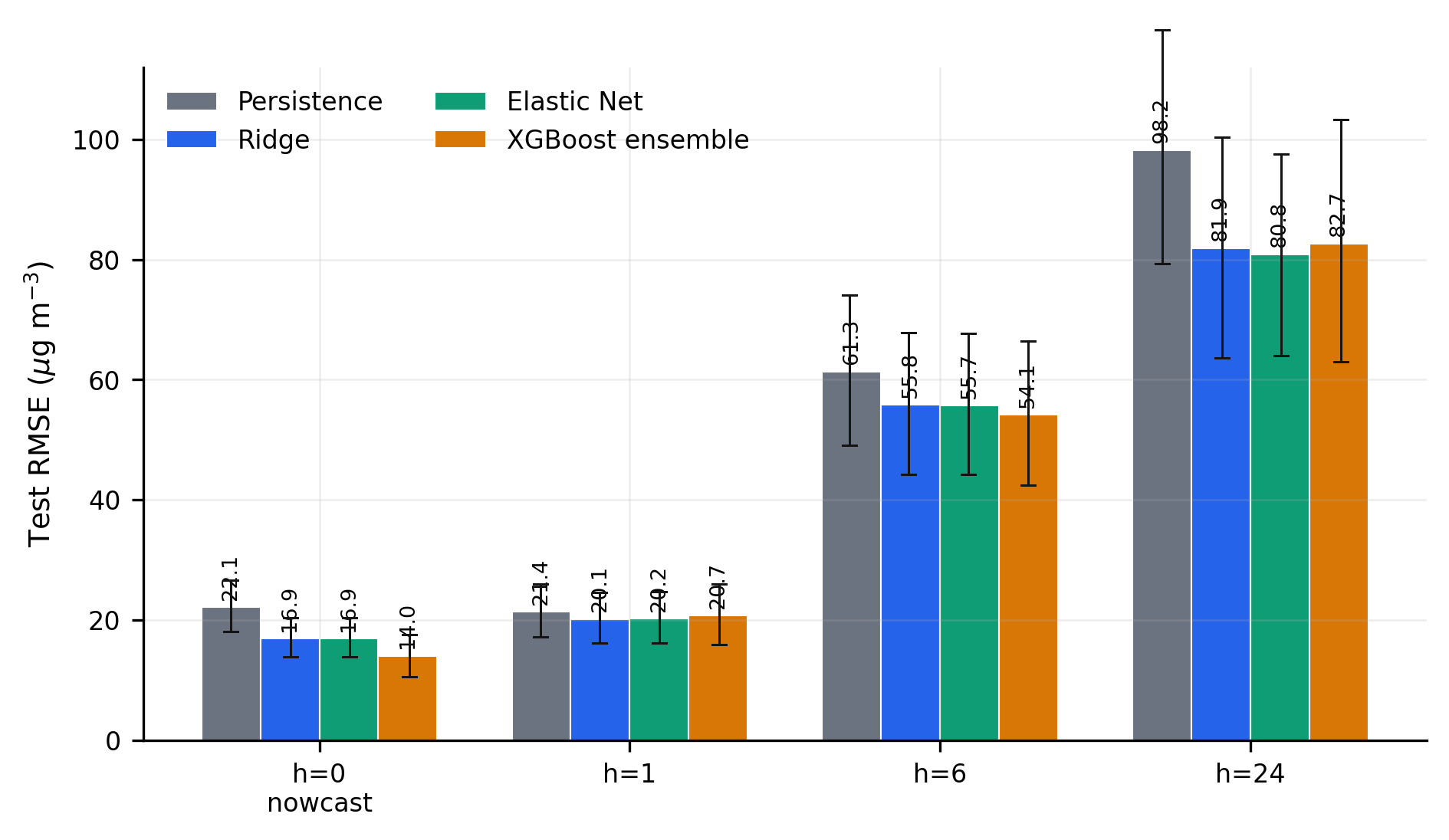}
    \caption{Test RMSE for same-hour estimation and strict one-, six-, and 24-hour
    forecasting. Error bars show 95\% moving-block bootstrap intervals; error increases
    with lead time, and the best-performing model changes across horizons.}
    \label{fig:primary}
\end{figure}

Table~\ref{tab:primary} and Figure~\ref{fig:primary} report final test performance.
The figure makes the central horizon pattern visible: absolute error rises sharply as the target
moves farther beyond the origin time, and the leading model is not constant across horizons.
At one hour, Ridge gives the lowest RMSE of 20.147, compared with 21.369 for persistence;
Elastic Net is nearly identical at 20.202, while XGBoost attains 20.735.
The short horizon therefore leaves relatively little room beyond the most recent PM$_{2.5}$
observation.

At six hours, XGBoost performs best with RMSE 54.131, improving on persistence by 11.75\%.
At 24 hours, Elastic Net gives the lowest RMSE of 80.829, a 17.65\% improvement over persistence,
followed by Ridge at 81.872 and XGBoost at 82.660.
The larger percentage improvement at 24 hours should be read together with the larger absolute
RMSE: the model adds information beyond persistence, but the underlying task is also substantially
more uncertain.

Same-hour estimation produces lower errors because additional synchronous measurements are
available.
The XGBoost ensemble reaches RMSE 14.027 and $R^2=0.977$, compared with RMSE 16.885 for Ridge
and 22.129 for the one-hour-lag persistence estimate.
This result describes a sensor-estimation setting rather than a forecast issued before the target
hour.
Its interpretation is therefore linked to the feature ablations below, which identify the
information responsible for the high apparent accuracy.
The corresponding $R^2$ values for the best point models are 0.953 at one hour, 0.660 at six
hours, and 0.242 at 24 hours (versus $-0.117$ for 24-hour persistence).
Thus, the longer-horizon result is not only a larger absolute error; it also represents a
substantial loss of explained variance.

\begin{table}[htbp]
\caption{Primary test-set performance. Improvement is relative to the persistence baseline
at the same horizon.}
\label{tab:primary}
\small
\resizebox{\textwidth}{!}{%
\begin{tabular}{@{}llrrrr@{}}
\toprule
\textbf{Task} & \textbf{Model} & \textbf{RMSE} & \textbf{MAE} & $\bm{R^2}$ &
\textbf{RMSE improvement (\%)}\\
\midrule
Nowcast, $h=0$ & Persistence & 22.129 & 11.067 & 0.943 & 0.00\\
Nowcast, $h=0$ & Ridge & 16.885 & 9.553 & 0.967 & 23.70\\
Nowcast, $h=0$ & Elastic Net (Lasso) & 16.920 & 9.445 & 0.967 & 23.54\\
Nowcast, $h=0$ & XGBoost ensemble & 14.027 & 7.384 & 0.977 & 36.61\\
Forecast, $h=1$ & Persistence & 21.369 & 10.721 & 0.947 & 0.00\\
Forecast, $h=1$ & Ridge & 20.147 & 10.544 & 0.953 & 5.72\\
Forecast, $h=1$ & Elastic Net (Lasso) & 20.202 & 10.457 & 0.953 & 5.46\\
Forecast, $h=1$ & XGBoost ensemble & 20.735 & 10.181 & 0.950 & 2.97\\
Forecast, $h=6$ & Persistence & 61.341 & 35.056 & 0.563 & 0.00\\
Forecast, $h=6$ & Ridge & 55.831 & 34.310 & 0.638 & 8.98\\
Forecast, $h=6$ & Elastic Net (Lasso) & 55.743 & 34.237 & 0.639 & 9.13\\
Forecast, $h=6$ & XGBoost ensemble & 54.131 & 31.876 & 0.660 & 11.75\\
Forecast, $h=24$ & Persistence & 98.151 & 65.172 & -0.117 & 0.00\\
Forecast, $h=24$ & Ridge & 81.872 & 56.729 & 0.223 & 16.59\\
Forecast, $h=24$ & Elastic Net (Lasso) & 80.829 & 57.127 & 0.242 & 17.65\\
Forecast, $h=24$ & XGBoost ensemble & 82.660 & 56.365 & 0.208 & 15.78\\
\bottomrule
\end{tabular}%
}
\end{table}

Because global averages can hide under-prediction during severe pollution episodes, we
performed a post-hoc subgroup evaluation on final test rows with PM$_{2.5}$ at or above the
raw-data 95th-percentile threshold of 242~$\mu$g/m$^3$.
The entries below report RMSE with $R^2$ in parentheses; the subgroup is defined after the
original chronological split and does not change model fitting.

\begin{table}[htbp]
\caption{Extreme-pollution test-set performance for target PM$_{2.5}\geq242~\mu$g/m$^3$.
Entries are RMSE ($R^2$).}
\label{tab:extreme}
\small
\begin{tabularx}{\textwidth}{lRRRRR}
\toprule
\textbf{Task} & $\bm{n}$ & \textbf{Persistence} & \textbf{Ridge} &
\textbf{Elastic Net (Lasso)} & \textbf{XGBoost}\\
\midrule
$h=0$ & 4111 & 45.98 (0.745) & 35.63 (0.847) & 35.95 (0.844) & 37.16 (0.834)\\
$h=1$ & 4098 & 44.38 (0.763) & 43.16 (0.776) & 43.41 (0.773) & 51.26 (0.684)\\
$h=6$ & 4091 & 130.37 (-1.043) & 139.51 (-1.339) & 138.85 (-1.317) & 139.07 (-1.324)\\
$h=24$ & 4069 & 184.76 (-3.087) & 226.77 (-5.156) & 218.26 (-4.703) & 230.92 (-5.384)\\
\bottomrule
\end{tabularx}
\end{table}

Ridge is best within the extreme subset at $h=0$ and $h=1$, whereas persistence is best at
$h=6$ and $h=24$; the global XGBoost advantage at six hours therefore does not transfer to
the highest-concentration rows.
The negative long-horizon $R^2$ values show that severe-event prediction remains difficult and
should not be inferred from global RMSE alone.

The paired moving-block intervals support the horizon-dependent comparison
(Table~\ref{tab:uncertainty_key}).
Ridge improves one-hour RMSE by 5.7\%, with a 95\% interval of 4.9--6.6\%.
The one-hour XGBoost improvement interval includes zero, whereas the best six- and 24-hour
models retain positive lower bounds.
Thus, selecting a more flexible learner at the shortest horizon is not supported by the paired
comparison, even though the point estimates differ modestly.

Random-seed variation is small relative to the between-horizon error differences: the XGBoost
RMSE mean and standard deviation are 14.12$\pm$0.11, 20.79$\pm$0.14, 54.23$\pm$0.19, and
82.83$\pm$0.54 for $h=0,1,6,24$, respectively.
The seed standard deviations are well below the difference between one- and six-hour errors,
so the broad horizon pattern does not depend on a particular ensemble draw.
The wider moving-block intervals at longer horizons instead reflect the changing difficulty and
serial dependence of the concentration process.

\begin{table}[htbp]
\caption{Paired moving-block uncertainty for the best point model at each strict
forecast horizon.}
\label{tab:uncertainty_key}
\small
\begin{tabularx}{\textwidth}{clRR}
\toprule
$\bm{h}$ & \textbf{Model} & \textbf{RMSE [95\% CI]} & \textbf{Improvement [95\% CI] (\%)}\\
\midrule
1 & Ridge & 20.15 [16.09, 24.52] & 5.7 [4.9, 6.6]\\
6 & XGBoost ensemble & 54.13 [42.43, 66.36] & 11.8 [8.8, 14.9]\\
24 & Elastic Net (Lasso) & 80.83 [63.94, 97.57] & 17.6 [10.8, 23.7]\\
\bottomrule
\end{tabularx}
\end{table}

\subsection{Feature-Group Contributions}

\begin{figure}[htbp]
    \centering
    \includegraphics[width=0.86\textwidth]{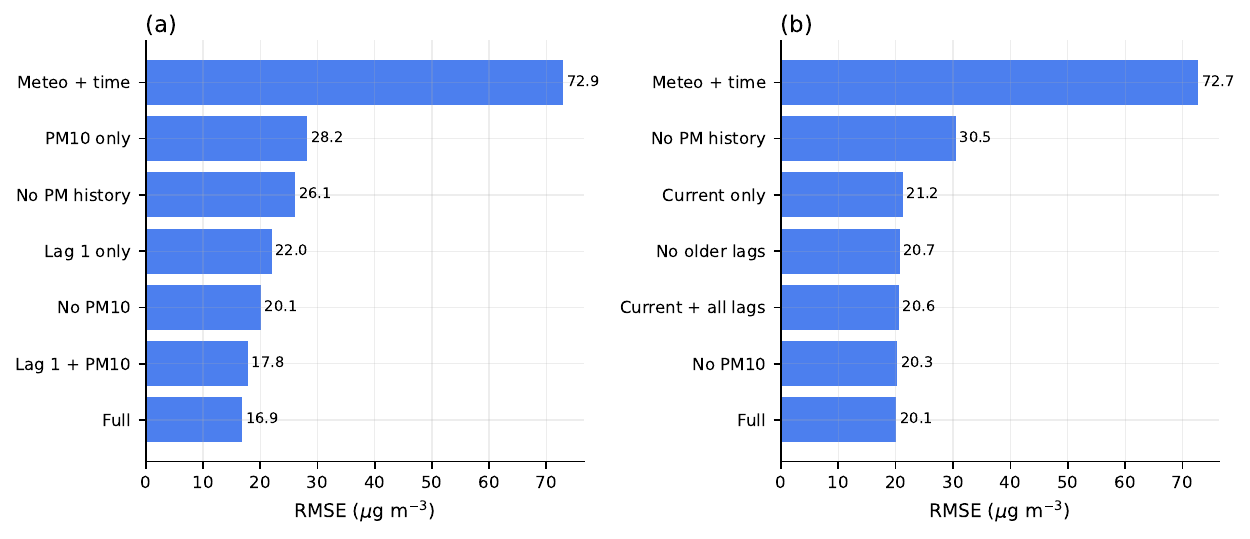}
    \caption{Ridge feature-group ablations for same-hour estimation ($h=0$, panel (a))
    and one-hour forecasting ($h=1$, panel (b)). Each bar reports test RMSE for a
    feature-set variant; the information hierarchy changes when the target is moved
    beyond the origin time.}
    \label{fig:ablation}
\end{figure}

The ablations in Table~\ref{tab:ablation_key} and Figure~\ref{fig:ablation} separate the
information behind the same-hour and one-hour results.
For $h=0$, a Ridge model containing only PM$_{2.5}$ lag 1 and synchronous PM$_{10}$ attains
RMSE 17.842, close to 16.885 for the full Ridge model.
Lag 1 alone gives 21.981 and PM$_{10}$ alone gives 28.182.
Removing PM$_{10}$ from the full feature set raises RMSE to 20.103.
Because PM$_{2.5}$ is physically a subset of PM$_{10}$, this synchronous input is not an
independent source of future information; it is a contemporaneous proxy that makes the $h=0$
task a sensor cross-check.
The same-hour score is therefore largely explained by recent target history and a strongly
correlated synchronous pollutant.

For the one-hour forecast, current PM$_{2.5}$ alone gives RMSE 21.235, close to persistence
and the full Ridge model at 20.147.
Removing PM$_{10}$ changes the full-model RMSE only to 20.269, while removing all current and
lagged PM$_{2.5}$ information raises RMSE to 30.471.
Co-pollutants and meteorology retain useful redundancy when the target sensor is unavailable,
but recent PM$_{2.5}$ remains the dominant short-horizon anchor.
The contrast between the two columns in Figure~\ref{fig:ablation} shows why a same-hour score
should not be used as a surrogate for forecast value.

\begin{table}[htbp]
\caption{Key Ridge ablations for same-hour estimation and one-hour forecasting (RMSE).}
\label{tab:ablation_key}
\small
\begin{tabularx}{\textwidth}{LRR}
\toprule
\textbf{Feature set }& $\bm{h=0}$ & $\bm{h=1}$\\
\midrule
Full & 16.885 & 20.147\\
Current PM$_{2.5}$ only & -- & 21.235\\
PM$_{2.5}$ lag 1 only & 21.981 & 33.610\\
Lag 1 + PM$_{10}$ & 17.842 & 29.027\\
No PM$_{10}$ & 20.103 & 20.269\\
No PM$_{2.5}$ history & 26.149 & 30.471\\
Meteorology + time & 72.917 & 72.708\\
PM$_{10}$ only & 28.182 & --\\
\multicolumn{3}{l}{\footnotesize Note: Current PM$_{2.5}$ is omitted at $h=0$ by design
to prevent a trivial identity mapping.}\\
\bottomrule
\end{tabularx}
\end{table}

To complement the Ridge ablations, we refitted the registered five-seed XGBoost configuration
and aggregated normalized gain-based feature importance into the same information groups.
Missingness indicators are included in the group to which their underlying variable belongs.
The values in Table~\ref{tab:xgb_importance} are mean percentages across seeds; they quantify
use by the fitted tree ensemble, not causal effects.

\begin{table}[htbp]
\caption{XGBoost group-level feature importance (mean normalized gain, \%).
Meteo. denotes meteorological variables.}
\label{tab:xgb_importance}
\small
\begin{tabularx}{\textwidth}{lRRRRRR}
\toprule
\textbf{$h$} & \textbf{PM$_{2.5}$ history} & \textbf{PM$_{10}$} &
\textbf{Other pollutants} & \textbf{Meteo.}\newline\textbf{/wind} &
\textbf{Calendar} & \textbf{Station}\\
\midrule
0  & 75.15 & 14.05 & 2.75 & 2.56 & 4.04 & 1.45\\
1  & 81.16 & 8.37  & 1.61 & 4.48 & 3.23 & 1.15\\
6  & 65.30 & 4.41  & 2.58 & 12.90 & 12.65 & 2.16\\
24 & 43.26 & 4.87  & 7.73 & 15.29 & 28.13 & 0.73\\
\bottomrule
\end{tabularx}
\end{table}

The XGBoost importance profile supports the Ridge ablation interpretation while adding a
nonlinear view: PM$_{2.5}$ history accounts for 75.15\% of gain at $h=0$, and synchronous
PM$_{10}$ contributes 14.05\%.
At $h=6$ and $h=24$, meteorology and calendar groups become more prominent, although recent
PM$_{2.5}$ history remains the largest group.
These values explain why the high $h=0$ score should not be interpreted as a general nonlinear
forecasting advantage.

\subsection{Temporal and Spatial Robustness}

Leave-one-station-out Ridge evaluation yields pooled RMSE values of 20.155, 55.891, and
82.076 at one, six, and 24 hours.
These values are close to the corresponding pooled-model results, although station-level
errors remain heterogeneous.
The result indicates that the pooled feature construction is not driven by a single station,
while also showing that a network-average score should not be treated as the expected error
at every location.

Split-conformal intervals have empirical coverage near their nominal levels for $h=0$ and
$h=1$, but coverage decreases and intervals widen at six and 24 hours.
Together with the moving-block results, this pattern indicates that longer lead times increase
both point error and predictive uncertainty.
For alert preparation, this favors communicating a forecast range or risk category in addition
to a point concentration, particularly at the six- and 24-hour horizons.

\section{Discussion}
\label{sec:discussion}

\subsection{Model Performance and Forecast Horizon}

The comparison shows that model rankings depend on forecast lead time.
Persistence is difficult to improve upon at one hour because the current PM$_{2.5}$ observation
already carries most of the short-term signal.
Ridge and Elastic Net nevertheless provide small, statistically stable improvements.
At six hours, nonlinear interactions represented by XGBoost become more useful, while at 24
hours the sparse linear model is competitive with or slightly better than the tree ensemble.
The same horizon dependence is visible in $R^2$: the best-model value falls from 0.953 at one
hour to 0.660 at six hours and 0.242 at 24 hours.
The decline means that longer-horizon point improvements coexist with much weaker explanatory
power, so RMSE improvement percentages should not be read as increased certainty.
The results therefore favor horizon-specific model selection rather than a single ranking
applied to every prediction task.

This pattern is consistent with a broader air-quality modeling literature in which neural,
convolutional, recurrent, and graph-based methods can be effective when the target horizon and
data structure justify the additional flexibility~\citep{li2017lstm,huang2018cnn,qi2019hybrid}.
The present comparison does not use identical input sets or splits to those studies, so RMSE
values should not be ranked across papers.
Its contribution is instead a like-for-like comparison within one monitoring network: the same
target definition, chronological partitions, baselines, and feature construction are used at
all four horizons.
This makes the observed switch from Ridge at one hour to XGBoost at six hours and Elastic Net
at 24 hours more informative than a comparison of isolated scores from different data sets.

The same-hour task occupies a distinct monitoring role.
Its lower error reflects access to synchronous co-pollutants and very recent target history.
Such estimates can support sensor cross-checking, quality control, or short delays in the
PM$_{2.5}$ channel.
Forecasting, by contrast, measures the value of an actual lead time and should be assessed
against persistence at the same horizon.
Separating these tasks also prevents a monitoring-system redundancy result from being presented
as evidence that future pollution levels can be predicted with the same accuracy.

Together, the strict forecasts define distinct operational roles across lead times.
At one hour, the modest 5.7\% improvement over persistence favors simple regularized updates
when low latency and stability matter.
The six-hour XGBoost improvement of 11.8\%, supported by a positive paired moving-block interval,
provides the strongest evidence for forecast-assisted alert preparation.
At 24 hours, the 17.6\% Elastic Net improvement can inform daily monitoring plans and resource
scheduling, but the wider error interval and lower conformal coverage call for caution in
threshold-based warnings.
The practical question is therefore not only which model has the smallest error, but also which
lead time is long enough to support a decision and short enough to retain actionable accuracy.

\subsection{Information Hierarchy and Monitoring Implications}

The ablations clarify how the role of each feature group changes across tasks.
Synchronous PM$_{10}$ is central to same-hour estimation, but its marginal contribution is much
smaller in the one-hour forecast once current PM$_{2.5}$ is available.
Meteorological and temporal variables become more relevant at longer horizons, although
meteorology and time alone remain weak.
The XGBoost gain profile in Table~\ref{tab:xgb_importance} gives the same qualitative hierarchy:
PM$_{2.5}$ history dominates at $h=0$ and 1, while calendar and meteorology account for a larger
share at $h=24$.
These patterns are consistent with short-term concentration persistence followed by an increasing
contribution from dispersion and temporal conditions as lead time grows.

The no-PM$_{2.5}$-history setting also provides a practical sensor-outage scenario.
Co-pollutants, meteorology, and calendar information reduce error well below a meteorology-only
model, but they do not recover the performance of a functioning PM$_{2.5}$ channel.
This is consistent with the sensor-monitoring literature: auxiliary sensors can strengthen
coverage and quality assurance, but calibration drift, pollutant-specific response, and local
environmental effects limit their role as replacements for reference measurements
~\citep{kumar2015rise,morawska2018applications,karagulian2019review,malings2019calibration}.

The distinction matters for a deployed network.
A same-hour model could flag disagreement between a PM$_{2.5}$ channel and its correlated
co-pollutant context, prompting inspection or temporary estimation.
Its output should be labeled as an estimate conditional on synchronous measurements, not as
an independently observed concentration.
Conversely, a strict forecast can be issued before the target hour but should account for the
uncertainty of its weather inputs and for the decline in calibration observed at longer horizons.
This division of roles supports a layered workflow in which direct sensing, sensor
cross-checking, and forecast guidance complement one another.

Such a workflow can use different model characteristics at different stages.
A regularized linear model is inexpensive to refit and its coefficients provide a direct audit
trail, which is useful for routine one-hour updates and for detecting a sudden change in the
relationship between the target and co-pollutants.
A tree ensemble can be reserved for the six-hour horizon, where nonlinear interactions provide
the clearest improvement over persistence.
At 24 hours, the similarity among Ridge, Elastic Net, and XGBoost errors suggests that data
availability and weather uncertainty may matter more than additional model complexity.
Operational selection can therefore consider latency, maintainability, and calibration alongside
point accuracy.

The uncertainty results also affect how forecasts should be communicated.
A positive average improvement does not imply that every pollution episode is predicted
accurately, and the widening intervals at longer horizons show that an isolated point value can
convey excessive precision.
The extreme-pollution subgroup reinforces this caution: persistence outperforms all fitted
models at $h=6$ and $h=24$, and all three strict-forecast models have negative $R^2$ values
in that subgroup.
Six-hour forecasts are most naturally used to support preparation and escalation decisions,
while 24-hour forecasts are better suited to planning and resource scheduling.
Threshold-based public warnings would require a separate evaluation of event detection, false
alarms, and interval calibration around the relevant concentration standards.

\subsection{Scope and Further Study}

The data represent one city and one monitoring network, so the numerical results should be
transferred to other locations only after external validation.
The leave-one-station-out analysis is informative about within-network spatial heterogeneity,
but it does not replace evaluation in another city, climate regime, or monitoring technology.
In addition, the meteorological predictors at origin time are observed measurements; a deployed
24-hour system would require forecast-weather inputs.
Replacing persistence-like origin observations with ECMWF, GFS, or another
numerical-weather-prediction product would introduce forecast error before the pollutant model
is applied.
That error could reduce accuracy and change the relative importance of meteorological variables
versus pollutant history, so weather-forecast substitution should be evaluated as a separate
operational experiment rather than inferred from the present retrospective results.

The lower conformal coverage at longer horizons further motivates interval methods designed
explicitly for dependent observations, including EnbPI and adaptive conformal inference
~\citep{xu2021enbpi,gibbs2021adaptive}.
Future work can examine weather-forecast substitution, cross-city transfer, and calibrated
probabilistic models within the same information-availability framework.
These extensions would preserve the key practical discipline of the present study: define the
forecast origin first, then evaluate the information and model that would genuinely be available
at that origin.

\section{Conclusions}
\label{sec:conclusion}

This study compared same-hour PM$_{2.5}$ estimation with strict one-, six-, and 24-hour
forecasting in Beijing.
The main findings are:
\begin{enumerate}
    \item Same-hour estimation achieves the lowest numerical error, but its accuracy is driven
          mainly by recent PM$_{2.5}$ history and synchronous PM$_{10}$.
          PM$_{10}$ is a strong contemporaneous proxy for the $h=0$ task, and this result does
          not demonstrate future forecasting skill.
    \item Persistence is a strong one-hour baseline; XGBoost gives the clearest improvement at
          six hours, while sparse linear models remain competitive at 24 hours.
    \item Paired temporal resampling, repeated seeds, and leave-one-station-out evaluation
          preserve the main horizon-dependent pattern while revealing greater uncertainty at
          longer lead times.
    \item The resulting monitoring roles are tiered by lead time: same-hour estimates support
          sensor cross-checking, one-hour models support routine updates, six-hour forecasts
          support alert preparation, and 24-hour forecasts support daily planning with explicit
          attention to uncertainty.
\end{enumerate}

Overall, information availability and forecast horizon determine both model rankings and the
value of individual feature groups.
A multi-horizon comparison therefore provides a more useful basis for PM$_{2.5}$ monitoring
and forecasting than a single same-hour accuracy score.
The resulting evidence supports sensor cross-checking at $h=0$, routine updates at one hour,
alert preparation at six hours, and uncertainty-aware daily planning at 24 hours.


\section*{Data Availability Statement}
The Beijing Multi-Site Air-Quality Data Set is publicly available from the UCI Machine
Learning Repository~\citep{ucibeijing2017}.
The accompanying code records source-file hashes, package versions, split boundaries, model
parameters, and the predictions used to construct the tables and figures.
Automated tests cover station isolation, clock-aligned lags, target shifting, rolling-window
exclusion of the current target, and timestamp-disjoint partitions.

\section*{Author Contributions}
Conceptualization, Y.W. and R.D.; methodology, Y.W.; software, Y.W.; validation, Y.W. and
R.D.; formal analysis, Y.W.; data curation, Y.W.; writing--original draft preparation, Y.W.;
writing--review and editing, Y.W. and R.D.; visualization, Y.W.; supervision, R.D.
All authors have read and agreed to the published version of the manuscript.

\section*{Funding}
This research received no external funding.

\section*{Declaration of Competing Interest}
The authors declare no conflict of interest.

\section*{Acknowledgements}
The authors thank the UCI Machine Learning Repository for providing the Beijing Multi-Site
Air-Quality Data Set.

\section*{Declaration of Generative AI and AI-Assisted Technologies in the Manuscript Preparation Process}
During the preparation of this work, the authors used Codex (provided by OpenAI)
for text editing, formatting, and manuscript organization.
After using this tool/service, the authors reviewed and edited the content as needed and take
full responsibility for the content of the publication.


\bibliographystyle{plainnat}
\bibliography{references}

\end{document}